\documentclass[aps,prl,twocolumn,superscriptaddress,showpacs,amsmath,amssymb]{revtex4}

\usepackage{graphicx}
\usepackage{amsmath}

\newcommand*{\cN}{{\cal N}}
\newcommand*{\tr}{\mbox{tr}\, }

\begin{document}
\title{Weak Coulomb blockade effect in quantum dots}

\author{Piet W.\ Brouwer}
\affiliation{Laboratory of Atomic and Solid State Physics, Cornell
University, Ithaca, New York 14853-2501}

\author{Austen Lamacraft}
\affiliation{Department of Physics, Princeton University, Princeton, NJ
  08544, USA}

\author{Karsten Flensberg}

\affiliation{Nano-Science Center, Niels Bohr Institute,
Universitetsparken 5, 2100 Copenhagen, Denmark}

\date{\today}

\begin{abstract}
We develop the general non-equilibrium theory of transport through
a quantum dot, including Coulomb Blockade effects via a $1/N$
expansion, where $N$ is the number of scattering channels. At
lowest order we recover the Landauer formula for the current plus
a self-consistent equation for the dot potential. We obtain the
leading corrections and compare with earlier approaches.
Finally, we show that to leading and next leading order in $1/N$
there is no interaction correction to the weak localization, in
contrast to previous theories, but consistent with experiments
by Huibers \textit{et al.}\ [Phys. Rev. Lett. \textbf{81}, 1917
(1998)], where $N=4$.
\end{abstract}

\maketitle

The Landauer formula relates the two-terminal conductance of a
mesoscopic conductor to its quantum mechanical scattering matrix
\cite{landauer}. This formula, and its extensions to multiterminal
geometries and other transport properties, has become one of the
cornerstones of the theory of quantum transport. However, because
the Landauer formula relies on the single-particle scattering
matrix, effects of electron-electron interaction are neglected.

In a chaotic quantum dot
the dominant part of the interaction is\cite{kurland,abg}
\begin{equation}
  H_c=  E_c(\hat N_{\mathrm{d}}-{\cal N})^2,
  \label{eq:Hee}
\end{equation}
where $E_c=e^2/2 C$ is the one-electron charging energy in terms
of the dot's capacitance $C$, $\hat N_{\mathrm{d}}$ is the number
of electrons on the quantum dot, and ${\cal N}$ is an offset
determined by the voltages on nearby gates. The simplest way to
include this interaction into the scattering formalism is to
replace $H_c$ by a self-consistent potential
\begin{equation} \label{mean_field}
   V_{\mathrm{d}}(t)=-\frac{e}{C}\left(
  \langle\hat N_{\mathrm{d}}(t)\rangle-\cN \right).
\end{equation}
where the expectation value of the number of electrons $\langle
\hat N_{\mathrm{d}}\rangle$ is computed in the presence of the
potential $V_{\mathrm{d}}$.
The self-consistent potential is of particular interest for
nonlinear and time-dependent transport \cite{btp}. It is generally
believed that Eq.~(\ref{mean_field}) is valid when the charge on the
dot fluctuates freely, either due to large
contact conductance or high temperatures, in which cases Coulomb
blockade is suppressed \cite{abg}. However, in this limit no
systematic evaluation of the corrections to the quantum transport
has been given for general non-equilibrium conditions. In this
Letter, we develop a theory of non-equilibrium transport that
reproduces the Landauer formula with the self-consistent potential
at lowest order and admits corrections as a power series in the inverse of the number of scattering channels $N$.

The situation we study is thus referred to as `weak Coulomb
blockade' \cite{abg}, as a hard Coulomb gap in the current-voltage
characteristic is not present. 
Historically, the influence of quantum charge fluctuations
on the transport properties of mesoscopic
junctions first focused on the coupling to the `electromagnetic environment' of an electron moving
through a single junction~\cite{SCT_book}. This situation was recently extended
to a single open contact in large-$N$ limit, similar in spirit to
the present paper, but restricted to a point scatterer
\cite{gz1}. To deal with the strong quantum charge fluctuations
of almost fully transmitting point contacts,
Flensberg~\cite{flensberg} and Matveev~\cite{matveev} introduced a
bosonized description (see also Ref.~\onlinecite{furusaki}). The
weak Coulomb blockade situation has also been addressed starting
from the tunnel junction limit, using the phase as variable
\cite{tunnel}, and good agreement with experiments on large
conductance metallic systems has been achieved \cite{golubevetal}.
Recently, the technique was extended to junctions of arbitrary
transparency \cite{nazarov}. A crucial simplification of all these
approaches was that once an electron enters the dot, it never
coherently returns to the leads. This was assumed to be
satisfactory for large dots where the typical dwell time $\tau_{\rm d}$ by far
exceeds the time scale set by inelastic processes.

\begin{figure}
\centerline{\includegraphics[width=0.45\textwidth]{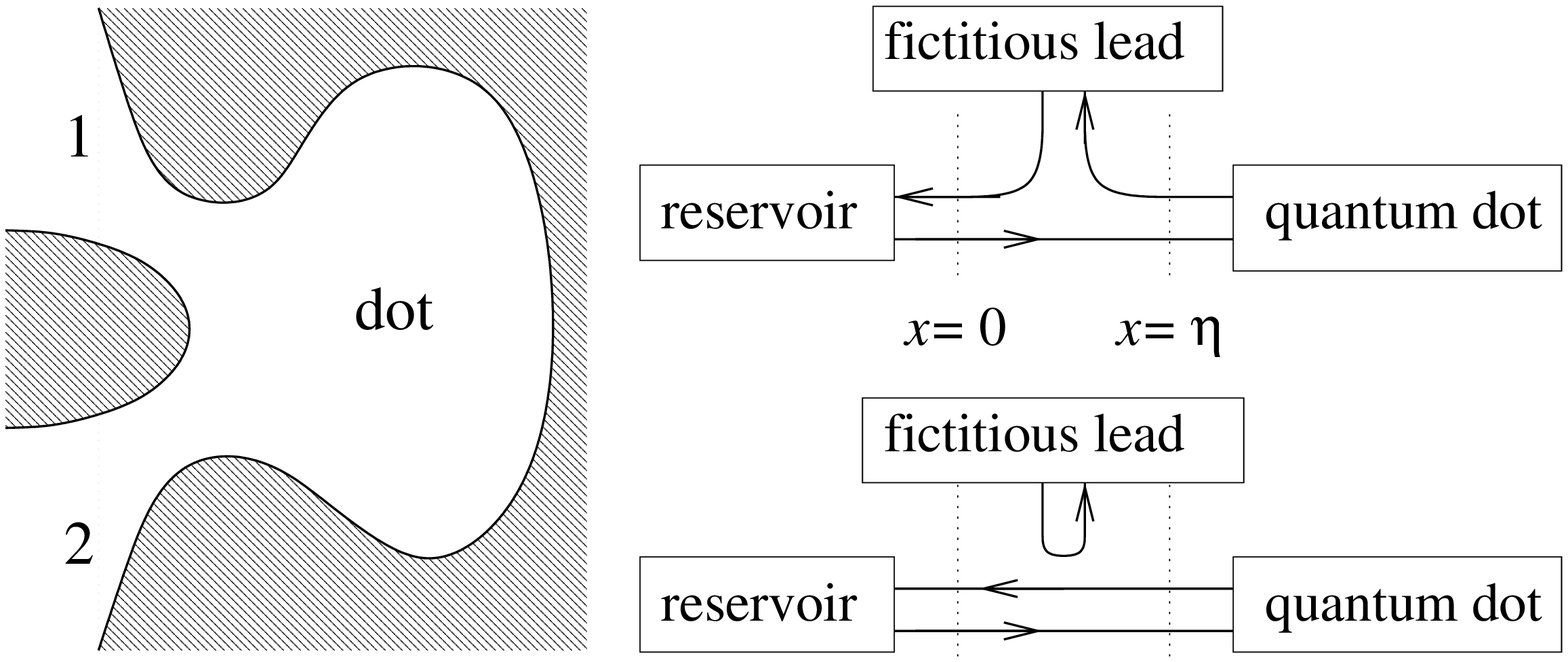}}
\caption{\label{fig:geometry} Left: Schematic drawing of a quantum
dot coupled to source and drain leads via point contacts.
Right, top: a fictitious lead is side-coupled to electrons leaving the
dot as a mathematical trick. The scattering Hamiltonian $H_1$,
when included to all
orders, decouples the fictitious leads, as illustrated in the
bottom right
panel.
}
\end{figure}

For the problem we consider here, the coherent return of the
electrons to the point contacts is crucial. Only then is transport
in the absence of interactions described by a unitary scattering
matrix $S$. For the coherent case, the interaction correction to
the dc conductance of an open quantum dot was first calculated in
the bosonization~\cite{abg,ba} and later in the phase
representation formalisms \cite{gz2}. The results of these two
approaches are strikingly different.
Our secondary aim in this Letter is to clarify this situation. In
this regard, we find that the result of Refs.\
\cite{abg,ba}, which is a formal expansion in the scattering
matrix $S$, is not applicable for unitary $S$, despite claims to
the contrary. 
However, it is applicable for a subunitary
scattering matrix, which may serve as a phenomenological
description of a quantum dot with strong relaxation.
On the other hand, the $1/N$ corrections we
obtain here are fully consistent with the analysis of
Ref.~\cite{gz2}. Furthermore, whereas Ref.\ \cite{gz2}
intended to describe the case of metallic quantum dots, {\em
i.e.}, neglecting quantum corrections to the non-interacting
conductance, we find that the same expression for the conductance
also describes the interaction correction to quantum interference
corrections. Thus our theory applies to experiments on the weak
localization corrections to conductance~\cite{huibers}.

We now turn to a description of our calculation. We consider a
quantum dot coupled to source and drain leads through point
contacts with $N_1$ and $N_2$ channels respectively ($N_1+N_2=N$),
see Fig.\ \ref{fig:geometry}. The starting point of our
calculation is the one-dimensional description of the electrons in
the point contacts \cite{flensberg,matveev}
\begin{eqnarray} \label{pc_H}
   H_0 &=&
  i v_F \sum_{j=1}^{N} \int dx
  \left[ \hat \psi_{Lj}^{\dagger}(x)
  \frac{\partial}{\partial x} \hat \psi_{Lj}^{\vphantom{\dagger}}(x)
  \right. \nonumber \\ && \left. \mbox{}
  - \hat \psi_{Rj}^{\dagger}(x)
  \frac{\partial}{\partial x} \hat \psi_{Rj}^{\vphantom{\dagger}}(x)
  \right]
  + H_c.
  \label{eq:H1d}
\end{eqnarray}
Here $v_F$ is the Fermi velocity (differing velocities in each channel may be removed by rescaling), $\psi_{Lj}(x)$ and $\psi_{Rj}$ describe left
and right moving fermions in channel $j$,
$H_c$ is given by Eq.\ (\ref{eq:Hee}),
and we set $\hbar = 1$. The
lead-dot interface is taken at $x=0$, the region $x < 0$ corresponding
to the leads.
Following Ref.\ \onlinecite{ag}, we write the number of electrons
in the dot as
\begin{equation}
  \hat N_{\mathrm{d}}= N_0 -
  \int_{-\infty}^{0} dx\! :\! [\hat \psi_{Lj}^{\dagger}(x)
  \hat \psi_{Lj}^{\vphantom{\dagger}}(x)
  + \hat \psi_{Rj}^{\dagger}(x)
  \hat \psi_{Rj}^{\vphantom{\dagger}}(x)]\!:
  \label{eq:nsubst}
\end{equation}
where $N_0$, the combined number of electrons in the dot
and the leads, is taken to be a constant.

The fields $\psi_{L}^{{}}$ and $\psi_{R}^{{}}$ in Eq.\
(\ref{eq:H1d}) are not independent, since they are connected by
the dot. After the substitution (\ref{eq:nsubst}), which removes
interactions inside the dot, their relationship is described by
the scattering matrix
\begin{equation} \label{S_rel}
  \hat\psi_{Li}^{{}}(\eta,t)=\sum_j\int dt'\,
  S_{ij}(t,t')\psi_{R\,j}^{{}}(\eta,t'),
\end{equation}
where $\eta$ is a positive infinitesimal. (Scattering from a
quantum dot with time-dependent potentials is described by a
scattering matrix with two time arguments, see {\em e.g.,} Ref.\
\cite{polianski}.)

Instead of solving the interacting Hamiltonian (\ref{eq:H1d}) with
the boundary condition (\ref{S_rel}), we reformulate the problem
using a trick that allows a treatment with standard perturbative
techniques. Our method is similar in spirit to that of Refs.\
\cite{abg,ba,ag}. As shown in Fig.~1, we couple a fictitious
reservoir
to the left-moving electrons in the point contact.
Then the problem reduces to the exactly solvable theory of
Flensberg and Matveev, where electrons that enter the quantum dot
never return coherently. We restore the original model by adding
a term
\begin{equation}
  H_1 = 2 v_F
  \sum_{j=1}^{N} \left( \hat \psi_{Lj}^{\dagger}(0)
  \hat \psi_{Lj}^{\vphantom{\dagger}}(\eta) +
  \hat \psi_{Lj}^{\dagger}(\eta)
  \hat \psi_{Lj}^{\vphantom{\dagger}}(0) \right),
  \label{coupH}
\end{equation}
to the one-dimensional Hamiltonian Eq.~(\ref{eq:H1d}), thus
shunting the fictitious leads and regaining coherent scattering by
the quantum dot.

We proceed by evaluating the current in perturbation theory in
$H_1$ using the Keldysh formalism
and Eq.~(\ref{S_rel}),
\begin{equation}
  I(t) = \left\langle T_{\mathrm c} \hat I(t) e^{-i \int_{\mathrm c} dt' \hat H_1(t')}
  \right\rangle_0.
  \label{eq:keldysh}
\end{equation}
Here $T_{\mathrm c}$ denotes ordering along the Keldysh contour c
and the average is taken with respect to the Hamiltonian $H_0$.
The current $I$ is calculated as the weighted difference of the
currents in the two point contacts,
\begin{equation} \label{current}
\hat I = -e v_F \sum_j\Lambda_{jj}\left[
\hat\psi_{R\,j}^{{\dagger}}(0)
  \hat \psi_{R\,j}^{\vphantom{\dagger}}(0) \!-\!
  \hat\psi_{L\,j}^{{\dagger}}(0) \hat \psi_{L\,j}^{\vphantom{\dagger}}(0)
  \right],
\end{equation}
where $\Lambda_{ij} = \delta_{ij} (N_2/N)$ for $j=1,\ldots,N_1$ and
$\Lambda_{ij} = -\delta_{ij} (N_1/N)$ for $j=N_1+1,\ldots,N$.

For the perturbative evaluation of Eq.\ (\ref{eq:keldysh}) we require the
correlation function~\cite{abg}
\begin{align}\label{correl}
   &\Big\langle T_c \prod_{p,q} \hat\psi^{\dagger}_{L\,i_p}(t_p)\hat\psi^{\vphantom{\dagger}}_{L\,l_p}(s_p)
   \hat\psi^{\vphantom{\dagger}}_{R\,j_q}(t'_q)\hat\psi^{\dagger} _{R\,k_q}(s_q')\Big\rangle_0
   =\nonumber\\
& \Big\langle \cdots \Big\rangle_{\mathrm{ni}}\times
  \prod_{p,q}\left(\frac{f(t_p,t_q')f(s_p,s_q')}{f(t_p,s_q')f(s_p,t_q')}\right)^{1/N},
\end{align}
(here all fields are taken at $x=0$) where $\langle\cdots\rangle_{\mathrm{ni}}$ represents the corresponding correlation function of the non-interacting theory (with $E_c \to 0$) and
\begin{align}\label{ln_f}
  \ln f(t,t') &= \mbox{const.}
      -\frac{E_c N }{\pi}\int_0^{\infty} d\zeta \,e^{-E_cN\zeta/ \pi }\\
      &\quad\times \ln \frac{\sinh\left[ \pi T
      \left( t-t'+\zeta-i0^+\mathrm{sgn}_c(t) \right) \right]}{\sinh\left[ \pi T\zeta \right]}.
\nonumber
 \end{align}
Here $\mathrm{sgn}_c(t)$ is $\pm 1$ for the forward and backward
branches of the Keldysh contour respectively. Additional `anomalous' correlators containing $N$ $\hat\psi^{\vphantom{\dagger}}_{L}$ and $N$ $\hat\psi^{\dagger}_{R}$ (or vice versa) give oscillating contributions to the current. These appear at $N^{\mathrm{th}}$ order in perturbation theory, so they are not captured by $1/N$ methods~\cite{abg}.

\begin{widetext}
Expanding Eq.\ (\ref{eq:keldysh}) to second order in $H_1$ is now
a straightforward calculation. The result is
\begin{eqnarray}\label{abg_res}
  I(t) &=&  -\frac{e}{2\pi}\mathrm{tr} \Lambda \mu_R
  -\frac{ie v_F}{2}
  \int dt' ds' \mathrm{tr}\left[ \Lambda S(t,t')G^K_{R}(t',s') S^{\dagger}(s',t) \right]\\*
 &&- e v_F^2 \mathrm{Im}\int ds dt' ds'\, \mathrm{tr}\left[\Lambda G^K_{L}(s,t)S(t,t')G^K_{R}(t',s')
 S^{\dagger}(s',s)  \right]\sin [\kappa_0(s,t')] \left|\frac{f(t,t') f(s,s')}{f(t,s') f(s,t')} \right|^{1/N},\nonumber
\end{eqnarray}
\end{widetext}
where we wrote $ N \kappa_0(t,t') {\mathrm{sgn}}_{\mathrm c}(t) =\mathrm{Im}\ln f(t,t')$,
{\em i.e.,}
\begin{eqnarray}
  \kappa_0(t,t')
  &=& \frac{\pi}{N}\left( 1-e^{-E_cN(t'-t)/\pi} \right)\theta(t'-t).
  \label{eq:phi}
\end{eqnarray}%
In Eq.\ (\ref{abg_res}) the trace is over the scattering channels $j=1,\ldots,N$. The
Keldysh Green's function $G_{Rij}^{K}(t,s)$ is
\begin{equation}
  G_{R\,ij}^K(t,s) = -\frac{ T}{v_F}
  \delta_{ij}{\mathrm P}\,\frac{e^{-i \int_s^t d\sigma \mu_{Ri}(\sigma)}}
  {\sinh\left[ \pi T (t-s) \right]},
  \label{eq:GKR}
\end{equation}
where ``P'' denotes the Cauchy principal value and $\mu_{Ri}$ is
the (electro-)chemical potential of electrons entering the dot
through through the $i$th channel, $i=1, \ldots,N$.
With zero chemical potential for the left movers,
electroneutrality demands $\sum_{j} \mu_{Rj} = 0$.
A similar expression holds for $G^{\mathrm K}_{L}$, but without
the phase factor in the exponent.

The linear dc conductance of Refs.\ \onlinecite{abg,ba} is
recovered upon taking $\mu_{Ri} = -e V_{\mathrm{sd}}
\Lambda_{ii}$, $i=1,\ldots,N$, setting $S(t,t') = S(t-t')$, and
expanding Eq.\ (\ref{abg_res}) in the source-drain voltage
$V_{\mathrm{sd}}$.
Truncating the perturbation theory in $H_1$ at second order
corresponds to a formal expansion up to second order in the
scattering matrix $S$ --- which is precisely what was done in
Refs.\ \onlinecite{abg,ba}. However, as $S$ is unitary, we find no
justification for stopping here.
That this approach gives unphysical results may be 
readily seen by substituting into Eq.~(\ref{abg_res}) the scattering
matrix of a ballistic quantum wire with no backscattering, resulting in a spurious correction to the Landauer conductance of $e^2/h$ per spin channel.
\begin{figure}
\centerline{\includegraphics[width=.4\textwidth]{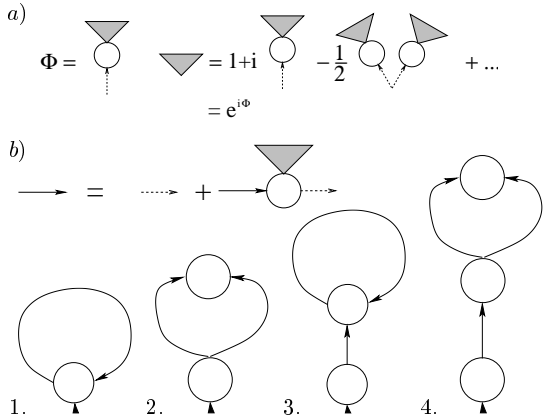}}
\caption{ a) At leading order in $1/N$, we retain only tree
diagrams. b) The one-loop diagrams\label{fig:diagrams}}
\end{figure}

In order to do a calculation to all orders in $H_1$, we need a
different principle to organize the perturbation theory.
In fact, the correlation function (\ref{correl}) shows that a
systematic expansion in $1/N$  is possible [keeping $E_CN$ in Eq.\
(\ref{eq:phi}) constant], while including $H_1$ to all orders.
The result is a sum over partitions into `bubbles' in standard
diagrammatic language. Each bubble is formally of order $N$, as it
involves a trace over the channel index. In order to be
non-vanishing, bubbles must be connected by `interaction lines'
$N^{-1}\ln f(t,t')$. Denoting these by dashed arrows, as in Fig.\
\ref{fig:diagrams}a, the arrowhead to indicate the unprimed time,
we see that at leading order we must have precisely one factor
$N^{-1} \ln f(t,t')$ for every bubble, except for the bubble that
contains the current operator. The result is a tree structure, as
shown in Fig.\ \ref{fig:diagrams}a. We thus find
\begin{eqnarray}
  \label{eq:I}
   I(t) &=& -\frac{e}{2 \pi} \mbox{tr}\, \Lambda \mu_R 
   \\
  && \nonumber \mbox{}  - \frac{ie v_F}{2}  \int dt'ds' \,\mathrm{tr}
  \Lambda S(t,t')\tilde G^K_R(t',s') S^{\dagger}(s',t).
\end{eqnarray}
Here $\tilde G^K_R(t',s')=G^K_R(t',s')e^{-i\tilde\phi(t',s')}$
where $\tilde\phi(t_1',t_2')$ satisfies the self-consistent
equation (see Fig.~\ref{fig:diagrams}a),
\begin{align}
  \tilde\phi(t',s') &= 2 \int_{-\infty}^{\infty}dt j_c(t)
  \left[ \kappa_0(t,t')-\kappa_0(t,s') \right],
  \label{eq:phiself}\\
  j_c(t) &= \frac{iv_F}{2} {\mathrm{tr}}\,
  \left[ G_{L}^{\mathrm K}(t,t) \vphantom{\int}
    \right. \nonumber \\ & \left. \mbox{} - \int dt_1' ds_1' S(t,t_1')\tilde G^K_R(t_1',s_1')
  S^{\dagger}(s_1',t)\right].
\end{align}
The current $j_c$ is the particle current leaving the dot due to
scattering inside.  The quantity $\tilde\phi(t',s')$ may be
thought of as the difference in phases accumulated by electrons
that enter the dot at times $t'$ and $s'$ in the presence of a
Coulomb potential. Indeed, writing
$\tilde\phi(t',s')=\int_{s'}^{t'}dt\, e V_{\mathrm d}(t)$ and
using Eq.\ (\ref{eq:phi}) for $\kappa_0$, the self-consistency
relation (\ref{eq:phiself}) takes the simple form
\begin{equation}
  C\partial_t V_{\mathrm d}(t)=e j_c(t)- (N e^2/2 \pi) V_{\mathrm d}(t),
\end{equation}
which is precisely the equation for the dot potential in the
self-consistent treatment of the interaction Hamiltonian
(\ref{eq:Hee}). Hence, evaluating the current at tree level, we
recover precisely the
Landauer formula Eq.~(\ref{eq:I}), including the approximation
(\ref{mean_field}) for the interaction~\cite{btp}.
The leading correction to the above self-consistent theory adds
one more `interaction line' to each tree, which results in
diagrams with one loop, and generates the RPA-like series in
Fig.~\ref{fig:diagrams}b. The summation of these leads to a
renormalization of the interaction line Eq.~(\ref{eq:phi}):
$\kappa(t,t')=\kappa_0(t,t')+ \int dt_1 \chi(t,t_1)
\kappa(t_1,t')$, where
\begin{widetext}
\begin{equation}\label{chi}
\nonumber
  \chi(t,t_1)=
    2\int dt_1'ds_1'  \,
  \frac{\delta j_c(t_1)}{\delta\tilde\phi(t_1',s_1')}
\left[\kappa_0(t,t_1')-\kappa_0(t,s_1') \right].
\end{equation}
%
The four one-loop diagrams are shown in Fig.~\ref{fig:diagrams}b,
where the solid interaction lines denote $\kappa(t,t')$
\cite{foot3}.
We thus find the leading interaction correction to the current is
\begin{eqnarray*}
  I_1(t)=  e v_F^2 \mbox{Im}\, \int dt_1 dt_1' ds_1' dt_2'ds_2' \kappa(t_1,t_2')
   \mathrm{tr} S(t_1,t_1')\tilde G^K_R(t_1',s_1') S^{\dagger}(s_1',t)\Lambda S(t,t_2')\tilde G^K_R(t_2',s_2') S^{\dagger}(s_2',t_1),
  \label{eq:I1}
\end{eqnarray*}
leading to our final result for the one-loop correction to the dc
conductance
\begin{eqnarray}
  G_1 =
  \frac{e^2}{8 \pi^2 T } \mbox{Im}\,
  \int d\varepsilon d\omega
   \frac{\tanh\left[(\varepsilon+\omega)/2 T\right]}{\cosh^2\left[\varepsilon/2 T\right]}
  \kappa(\omega)  \tr
  \left[\Lambda S(\varepsilon) \Lambda S^{\dagger}(\varepsilon + \omega)
  - \Lambda S(\varepsilon) S^{\dagger}(\varepsilon + \omega)
  S(\varepsilon) \Lambda S^{\dagger}(\varepsilon) \right].
  \label{gz_res}
\end{eqnarray}
\end{widetext}

The result (\ref{gz_res}) coincides with that found in
Ref.~\cite{gz2}. We note that it gives not only the leading
interaction correction to the semiclassical conductance, but also
to the quantum interference corrections. In particular, an
ensemble average of Eq.\ (\ref{gz_res}) predicts the absence of an
interaction correction to weak localization, which is in
disagreement with the conclusions of Ref.\ \onlinecite{ba}, but consistent with experiments by Huibers {\em et al.}, who find that weak localization is well described by the non-interacting theory, suitably modified for the presence of
a small amount of dephasing, down to the lowest temperatures available
\cite{huibers}.
There is a simple reason for this. The interaction correction
Eq.~(\ref{gz_res}) derives from a renormalized scattering matrix
$S'(\varepsilon)=S(\varepsilon)+\delta S(\varepsilon)$,
with~\cite{gz2,blf_long}
\begin{eqnarray}\label{deltaS}
\delta S(\varepsilon) &=&
  -i\int \frac{d\omega}{2\pi}\kappa(\omega)\left[
  \tanh\left[ \left( \varepsilon-\omega \right)/2T \right] S(\varepsilon-\omega)
  \right. \nonumber\\ &&\mbox{}  \left.  
   +   \tanh\left[ \left( \varepsilon+\omega \right)/2T \right]
  S(\varepsilon)S^{\dagger}(\varepsilon+\omega)S(\varepsilon) \right].\ \ \ \
\end{eqnarray}
This correction preserves, at each energy, the circular invariance
of the ensemble of scattering matrices that describes open dots.
Quantities like the weak localization correction, that do not
depend on correlations at different energies, are thus completely
unaffected by the replacement $S(\varepsilon)\to S'(\varepsilon)$. Note that this is a statment about the \emph{ensemble averaged conductance}, and the conductance of any particualar dot configuration will acquire a temperature-dependent renormalization of random sign.
We note that for non-ideal contacts with reflection matrices, $r_{{\rm c}j}$, $j=1,2$, Eq. (\ref{gz_res}) predicts the renormalization of the classical conductance $g\equiv g_1g_2/(g_1+g_2)$ 
\[ \delta g_j = - 2\tr r_{{\rm c}j}^{\vphantom{\dagger}} 
  r_{{\rm c}j}^{\dagger} ( 1 - r_{{\rm c}j}^{\vphantom{\dagger}} 
  r_{{\rm c}j}^{\dagger})\frac{1}{g} \ln \frac{E_{\rm c} g e^{1 + {\rm
  C}}}{2 \pi^2 T},\ \ \tau_{\mathrm d}^{-1}\ll T \ll g E_{\rm c}.  \]
$g_j=N_j-\tr r_{{\rm c}j}^{\vphantom{\dagger}} 
  r_{{\rm c}j}^{\dagger} $ is the conductance of each contact~\cite{gz1}.

We have presented a formalism that elucidates the relationship
between the conflicting theories of weak Coulomb blockade
\cite{ba,abg,gz2}. Our formalism provides systematic corrections
to the use of a self-consistent potential in the scattering
approach to quantum transport.

We would like to thank Igor Aleiner and Leonid Glazman for
discussion of the results. This work was supported by the NSF
under grant no.\ DMR 0334499 and by the Packard Foundation and by
the Danish Natural Science Reseach Council.

\end{document}